\documentclass[twocolumn]{aastex62}
\usepackage{amsmath}

\usepackage{makecell}

\usepackage{xcolor}

\graphicspath{{./}{figures/}}

\received{2019 October 29}
\accepted{2020 May 8, The Astrophysical Journal}

\shorttitle{VLA observations of high-z GRB hosts}
\shortauthors{Gatkine et al.}

\begin{document}

\title{New radio constraints on the obscured star formation rates of massive GRB hosts at redshifts $2-3.5$}


\correspondingauthor{Pradip Gatkine}
\email{pgatkine@astro.umd.edu}

\author[0000-0002-1955-2230]{Pradip Gatkine}
\affil{Dept. of Astronomy, University of Maryland, College Park, MD 20742, USA}

\author{Stuart Vogel}
\affiliation{Dept. of Astronomy, University of Maryland, College Park, MD 20742, USA}

\author[0000-0002-3158-6820]{Sylvain Veilleux}
\affiliation{Dept. of Astronomy, University of Maryland, College Park, MD 20742, USA}
\affiliation{Joint Space Science Institute, University of Maryland, College Park, MD 20742, USA}
\affiliation{Space Telescope Science Institute, Baltimore, MD 21218, USA}

\begin{abstract}
It is not clear whether gamma-ray bursts (GRBs) are unbiased tracers of cosmic star formation at $z > 2$. Since dusty starburst galaxies are significant contributors to the cosmic star formation at these redshifts, they should form a major part of the GRB host population. However, recent studies at $z \leq 2$ have shown that the majority of the star formation activity in GRB hosts is not obscured by dust. Here, we investigate the galaxy-scale dust obscuration in $z \sim 2-3.5$ GRB hosts pre-selected to have high-resolution, high signal-to-noise afterglow spectra in the rest-frame ultraviolet (UV) and thus relatively low line-of-sight dust obscuration. We present new deep VLA observations of four GRB hosts,
and compare the radio-based (upper limits on the) ``total" star formation rates (SFRs) to the ``unobscured" SFRs derived from fits to the optical-UV spectral energy distribution. The fraction of the total SFR that is obscured by dust in these galaxies is found to be $<$ 90\% in general, and $\la$ 50\%  for GRB 021004 in particular.
These observations suggest that $z \sim 2-3.5$ GRBs with UV-unobscured sightlines originate in star-forming galaxies with low overall 
dust obscuration, unlike the dusty starburst population.

\end{abstract}


\section{Introduction} \label{sec:intro}


Long gamma-ray bursts (GRBs) are bright bursts of gamma-rays followed by extremely luminous multi-wavelength afterglow, from the X-rays to the radio wavelengths. They have been shown to be associated with the collapse of massive stars \citep{hjorth2003very, stanek2003spectroscopic}. GRBs have been observed across the cosmic history, from $z \sim 0.01$ to $z \sim 8.2$ \citep{tanvir2009gamma, salvaterra2009grb, fynbo2000hubble}. 
These attributes make them a viable probe for tracing the star-formation history of the universe, especially at $z > 2$ where other probes are scarce.

However, the exact relation between GRB rates and cosmic star formation rate (SFR) is still an unsolved problem \citep{greiner2015gamma, schulze2015optically, perley2016swift_1, perley2016swift_2}. Various observations of  $z < 1.5$ GRB hosts have raised questions on whether GRBs can be used as unbiased tracers of star formation \citep{boissier2013method, perley2013population,  vergani2015long, schulze2015optically, perley2016swift_2}. Particularly, GRB hosts at $z < 1$ show a strong bias towards faint, low-mass ($\mathrm{M_{*} < 10^{10}~M_{\odot}}$), star-forming galaxies and lower metallicities (below solar metallicity) compared to other 
star-formation tracers, even after taking into account GRBs with high line-of-sight dust obscuration \citep{graham2013metal, perley2013population, kelly2014host, vergani2015long, japelj2016long, perley2016swift_2}. However, this bias appears to subside at $z > 2$ \citep{greiner2015gamma} since the mean metallicity of typical star-forming galaxies is below solar. A significant amount of star formation at these redshifts is contributed by dusty massive starbursts (submm-bright; see \cite{casey2014dusty} for a review). Thus, high-mass, (relatively) metal-rich, dusty galaxies with high star formation rates may form a significant fraction of the GRB host population at $z > 2$ \citep{perley2013population, greiner2016probing, perley2016swift_2}. 
 On the other hand, some previous studies indicate that GRB explosions may have a bias against dusty host galaxies based on the relatively stronger Ly-$\alpha$ emission of the hosts \citep{fynbo2003lyalpha} and the higher incidence of GRBs in the brightest regions in the galaxy compared to core-collapse supernovae \citep{fruchter2006long}.  To understand whether GRBs truly trace star formation at $z > 2$, it is important to measure the total SFR (i.e. dust-obscured + dust-unobscured).


Radio observations provide a probe of recent total star formation rate. In star-forming galaxies, the radio luminosity at frequencies below a few $\times$ 10 GHz is dominated by the synchrotron emission from relativistic electrons, previously accelerated by supernova remnants, propagating in the interstellar magnetic field \citep{condon1992radio}. The relativistic electrons probably have lifetimes $\leq$ 100 Myr, thus this component traces recent ($ <$ 100 Myr) star formation. 

There are about 100 GRB host observations at radio frequencies down to limits between $3 - 500$ $\mu$Jy (see \cite{greiner2016probing} for details). So far, there have been 19 cm-wave observations of GRB hosts at $z > 2$, out of which two were detections: GRB 080207A and GRB 090404  \citep{greiner2016probing, perley2013population, perley2015connecting, perley2016late}. However, none of these high-z GRBs have high-resolution, high-SNR afterglow spectra. 

GRBs with high-resolution afterglow spectra can be excellent test cases for examining the biases in GRB host population at high-z since a measure of the host metallicity may be derived from these spectra to help characterize the galaxy population traced by GRBs at $z > 2$. The availability of a high-resolution rest-frame UV spectrum of the GRB afterglow implies that the rest-frame UV is largely unobscured  (${A_{UV} \la}$ 2-3 mag). The radio observations of these GRB hosts may be used to find out whether this lack of obscuration is simply due to a clear line-of-sight or due to an overall lack of dust obscuration in the host galaxy. Dusty sightlines do not necessarily imply dusty host galaxies. This needs to be tested, especially in light of past cm-wave observations of \cite{hatsukade2012constraints} and \cite{perley2013population}, where the deep upper limits on the radio flux from the galaxy hosts of so-called `dark GRBs' (i.e. UV-dark afterglow due to high line-of-sight extinction) imply that the dark GRBs do not always occur in galaxies enshrouded by dust or in galaxies exhibiting extreme star formation rates (few $\times$ 100 $-$ 1000 $\mathrm{M_{\odot} yr^{-1}}$).

New radio-based SFR constraints are particularly needed for massive ($\mathrm{M_{*} \gtrsim 10^{10} M_{\odot}}$) GRB hosts at $z > 2$ since the massive star-forming galaxies at high-z are likely to be significantly dusty \citep{casey2014dusty, shapley2011physical}. One of our objectives is therefore to understand whether massive GRB hosts at $z > 2$ share this characteristic of typical massive star-forming galaxies at $z > 2$. 

This pre-selection of $z > 2$ GRB hosts based on high-resolution afterglow spectra is also useful to inform the total SFR of the GRB hosts in the CGM-GRB sample \citep{gatkine2019cgm}, particularly for the massive GRB hosts which are likely to have a substantial dust-obscured star formation component. The high-resolution spectra quantitatively trace the kinematics of the circumgalactic and interstellar media of the host. The total star formation (obscured $+$ unobscured) is a major driver of galactic outflows that feed the circumgalactic medium (CGM). Therefore, constraining the total SFR is necessary for studying the CGM-galaxy connection. 

In this paper, we report deep, late-time radio observations of four $z > 2$ GRB hosts with existing high-resolution afterglow spectra. The sample includes GRB 080810 which is the highest-redshift GRB host yet ($z = 3.35$) with deep radio observations. These results were obtained using Karl Jansky Very Large Array (VLA) in C-band ($4-8$ GHz). Section \ref{sec:data} describes the target selection, VLA observations, and analysis. In section \ref{sec:Total_SFR_estimate}, we derive the constraints on the radio-based SFRs and discuss the obscured fraction of the SFR in each GRB host individually. The implications of these results for dust obscuration in GRB hosts are discussed in Section \ref{sec: discussion} and the key conclusions are summarized in Section \ref{sec:summary}.


\section{Sample and observations} \label{sec:data}
\subsection{Sample Selection}
\label{subsec:sample}

The CGM-GRB sample is a sample of 27 $z > 2$ GRBs with high-resolution (resolving power $R$ $>$ 6000) and high signal-to-noise ratio (median SNR $\sim$ 10) afterglow spectra \citep{gatkine2019cgm}. None of these GRBs have previously reported late-time radio observations. A subset of these objects is selected by imposing various criteria. Only GRBs that occurred at least six years ago are considered to ensure that the radio flux contribution from the afterglow is minimal \citep{perley2015connecting}. From the remaining 17, only GRB hosts with existing $\mathrm{M_{\star}}$ measurements and $\mathrm{M_{\star} > 10^{9.5}~M_{\odot}}$ are selected since their UV-based SFR is expected to be most affected by dust obscuration. This resulted in a set of four GRB hosts: GRB 021004, GRB 080310, GRB 080810, and GRB 121024A. Further, the VLA observations of GRB 080810 reported here (at $z = 3.35$) make it the the highest-redshift GRB with a late-time radio observation of the host. Table \ref{tab:GRB_list} summarizes the sample and its key properties. 

\begin{deluxetable*}{ccccccccccc}[t]
\tablecaption{Summary of the VLA observations \label{tab:GRB_list}}
\tablehead{
\colhead{GRB\tablenotemark{a}} &
\colhead{$z$} &
\colhead{R.A.} &
\colhead{Dec.} & 
\colhead{Date} & 
\colhead{\makecell{$t_{int}$\\ (min)}} & 
\colhead{\makecell{Total $t_{int}$\\ (min)}} & 
\colhead{\makecell{3$\sigma$ Limit\\ ($\mu$Jy)}} & \colhead{\makecell{Beam size\\ ($''$)}} &
\colhead{\makecell{Flux/\\ bandpass}} &
\colhead{\makecell{Complex\\ gain}}}

\startdata
021004 & 2.323 & 00:26:54.68 & +18:55:41.6 & \makecell{2018 Dec 16\\ 2018 Dec 18} & \makecell{120.5\\ 150} & 270.5 & 3.0 & 3.7 $\times$ 4.5 & 3C48 & \makecell{J0010+1724}\\
080310 & 2.427 & 14:40:13.80 & $-$00:10:30.7 & \makecell{2018 Dec 04\\ 2018 Dec 11\\ 2018 Dec 15\\ 2018 Dec 18\\ 2018 Dec 24 }  & \makecell{90\\ 66\\ 90\\ 90\\ 66} & 402	& 6 & 3.2 $\times$ 4.0 & 3C286 & J1445+0958\\
080810 & 3.35 & 23:47:10.49 & $+$00:19:11.5 & \makecell{2018 Dec 09\\ 2018 Dec 22\\ 2019 Jan 05\\ 2019 Jan 10}  & \makecell{71\\ 135\\ 71\\ 66} & 343 & 3.8 & 3.7 $\times$ 4.9 & 3C48 & J2323-0317\\
121024A & 2.298 & 04:41:53.30  & $-$12:17:26.6  & \makecell{2018 Dec 17}  & \makecell{123}  & 123 & 12 & 3.9 $\times$ 5.6 & 3C138 & J0437-1844\\
\enddata

\tablenotetext{a}{All the observations were performed in the C-band ($4-8$ GHz) in C array configuration of the VLA}
\end{deluxetable*}

\subsection{VLA Observations} \label{sec:methods}
We performed the radio observations using the fully upgraded Karl G. Jansky Very Large array (VLA) using C-band receivers spanning 4 $-$ 8 GHz and with a central frequency of 6 GHz. We used 3-bit samplers to utilize the entire 4096 MHz bandwidth of the C band to maximize the continuum sensitivity. The dual polarization setup was used. The observations were conducted in the C array configuration during the months of December 2018 to January 2019 (program VLA 18B-312, PI: Gatkine). The integration time for each GRB host is listed in Table \ref{tab:GRB_list} (typical $\sim$ 4.5 hours). A nearby complex gain (amplitude and phase) calibrator was observed every 30 $-$ 40 minutes during any scheduling block and a standard flux calibrator was observed every hour. The 3-$\sigma$ rms and the synthesized beam size for each source are listed in Table \ref{tab:GRB_list}. 

The data reduction was carried out using the Common Astronomy Software Applications package (CASA) version 5.5.0. The standard {\fontfamily{qcr}\selectfont CASA} pipeline was used to flag and calibrate the observations. Imaging and deconvolution was performed using the {\fontfamily{qcr}\selectfont tclean} function in {\fontfamily{qcr}\selectfont CASA}. Natural weighting was employed while cleaning the measurement sets to maximize the continuum sensitivity. In the case of GRB 121024A, additional flagging was performed to clip the outlier visibilities and channels heavily affected with radio frequency interference. Further, self-calibration was performed to clean the image around a bright source at a separation of 6$^\prime$, a robust weighting was employed, and a  multi-term multi-frequency synthesis (mtmfs, with 2 terms) deconvolver was used to account for spectral index gradient in the much brighter contaminating source.    

The synthesized beam size for C-configuration observations is significantly coarser (beam size $\sim$ 4\arcsec) than the angular extent of the galaxy (1 kpc translates to $\sim$ 0.1\arcsec\ at $z \sim 2.5$). Therefore, the host galaxies are unresolved and can be treated as point sources here. The 1$\sigma$ flux-density level was derived by sampling a blank region spanning $\sim$100 $\times$ synthesized beam area around the target.  

The maps for GRB 021004 and GRB 080810 have rms sensitivities close to that predicted by the VLA noise calculator.  However, GRB 121024A and GRB 080310 had particularly bright sources near the half-power response of the primary beam.  At this location in the primary beam, the amplitude response is variable owing to antenna pointing errors, which result in amplitude gain errors in the visibilities that are a function of field position in addition to antenna, frequency, and time. Standard self-calibration does not work well if there are position-dependent errors; antenna pointing errors limited the dynamic range of the maps for GRB 080310 and especially GRB 121024A, and consequently our sensitivity for these objects.

\begin{figure*}
\centering
\includegraphics[width=0.9\textwidth]{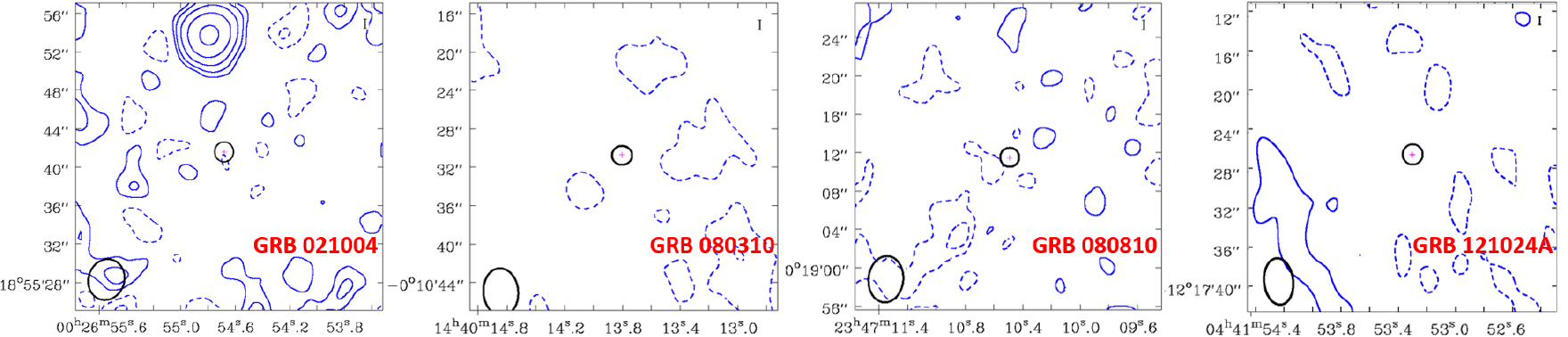}
\figcaption{\label{fig:VLA_postage} 
 Contour maps of the radio flux density in  $30\arcsec\ \times 30\arcsec$ fields centered on the four GRBs of our sample. The location of the GRB and 2\arcsec\ error circle are marked as red crosses and black circles, respectively. The synthesized beam is shown in the bottom left corner. The contours are marked as -12, -6, -3, -1.5, 1.5, 3, 6, 12 $\times \sigma$ with negative values marked as dotted contours. None of the GRB hosts are detected at the 3$\sigma$ level.}
\end{figure*}

\section{Radio- and UV-based  SFR}\label{sec:Total_SFR_estimate}

\subsection{Radio-based SFR}\label{sec:radio_SFR}

\begin{figure}
\centering
\includegraphics[width=0.425\textwidth]{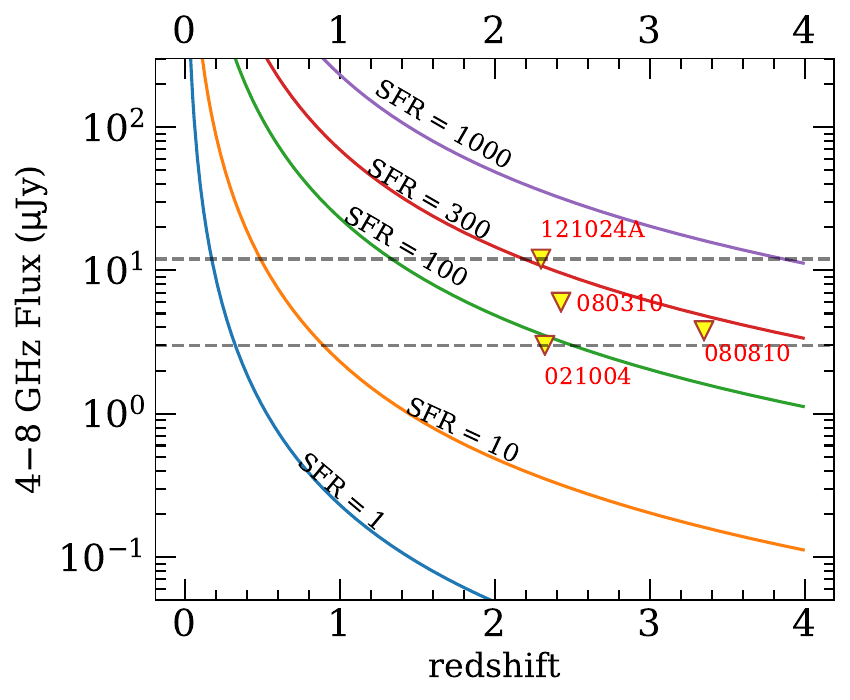}
\figcaption{\label{fig:Flux_vs_z} 
Curves showing the radio flux density averaged over $4-8$ GHz for various star formation rates ($\mathrm{M_{\odot}/yr}$) over a redshift range $z \sim 0 - 4$ using a spectral index of $\alpha = 0.7$. The 3$\sigma$ upper limits of various GRBs are shown with downward triangles. The horizontal dotted lines are drawn to guide the eye.}
\end{figure}

As described in Section \ref{sec:intro}, the radio continuum at frequencies below a few $\times$ 10 GHz traces the total (i.e. dust-obscured $+$ dust-unobscured) star formation activity in the last 100 Myr \citep{condon1992radio}. The radio-far-IR relation for star-forming galaxies which quantifies the radio-SFR relation is shown to hold true at intermediate and high redshifts \citep{sargent2010vla}. On the other hand, the UV/optical light (including the emission lines) primarily probes the portion of the SFR that is not significantly obscured by dust (i.e. dust-unobscured SFR) even with dust attenuation included in the modeling (see the example of GRB 100621A in \citealt{stanway2014radio}). Thus, a significant discrepancy between the UV-based and radio-based SFR measures would imply the presence of substantial dust obscuration within the galaxy.
\noindent In the discussion below, we use the following naming system:\\
\newline $\mathrm{SFR_{total}}$: Radio-based total SFR
\newline $\mathrm{SFR_{unobscured}}$: UV-based unobscured SFR (without dust correction),  
\newline $\mathrm{SFR_{obscured}}$: the portion of SFR that is obscured due to the dust (= $\mathrm{SFR_{total} - SFR_{unobscured}}$).\\


Here, we observe the GRB hosts in C-band ($4-8$ GHz) at $z \sim 2-3.5$, thus we are sensitive to $\nu_{rest} = 25 \pm 10$ GHz. The rest-frame radio luminosity is produced by three mechanisms: non-thermal synchrotron emission ($\epsilon_{1}$), free-free emission ($\epsilon_{2}$), and thermal emission from dust ($\epsilon_{3}$), as shown in  \cite{yun2002radio}. The thermal dust component is insignificant ($<$ 1\%) at the frequencies of interest. The radio-SFR relation for star-forming galaxies \citep{yun2002radio} is thus given by: 

\begin{mathletters}
\begin{eqnarray} \label{eqn:M_CGM}
    S(\nu_{obs}) & = & \Big( \epsilon_{1} + \epsilon_{2} + \epsilon_{3} \Big) \times \frac{(1+z)SFR}{D_{L}^{2}}
\end{eqnarray}
\end{mathletters}

\noindent where,

\noindent
$\epsilon_{1}  =  25f_{nth}\nu_{0}^{-\alpha}$\\
$\epsilon_{2} = 0.71\nu_{0}^{-0.1}$ \\
$\epsilon_{3} = 1.3 \times 10^{-6}  \frac{\nu_{0}^{3}[1 - e^{-(\nu_{0}/2000)^{\beta}}]}{e^{0.048\nu_{0}/T_{d} -1}}$. \\

Here, the symbols $\epsilon_{1}$, $\epsilon_{2}$, and $\epsilon_{3}$ represent 
the contributions from non-thermal synchrotron, free-free, and dust thermal emission respectively. $D_{L}$ is luminosity distance in Mpc, SFR is star formation rate in $\mathrm{M_{\odot} yr^{-1}}$, $\nu_{0}$ is rest-frame frequency in GHz, $f_{nth}$ is the scaling factor, $\alpha$ is the synchrotron spectral index, $T_{d}$ is the dust temperature in K, and $\beta$ is the dust emissivity. For the typical values of $T_{d}$ ($\sim$ 60 K) and $\beta$ (1.35), the dust emission is insignificant for $\nu_{\rm rest} \sim 25$ GHz. hence, we neglect this term. The non-thermal synchrotron emission is the most dominant contributor in the given frequency range. Since we do not have a robust measurement of the actual spectral index for any of our objects, we assume a canonical average value of $\alpha = -0.7$. Past literature has used values ranging from $-0.6$ to $-0.75$ \citep{hatsukade2012constraints, perley2013population, perley2015connecting, stanway2014radio, greiner2016probing}. This range of $\alpha$ affects the radio luminosity by 25\%. 
This equation assumes a Salpeter initial mass function (IMF). Due to various assumptions in the  calibration of radio-based SFRs, it is subject to a systematic uncertainty of about a factor of $\sim$2 \citep{yun2002radio, bell2003estimating, murphy2011calibrating}.  

Figure \ref{fig:Flux_vs_z} shows the observed flux densities averaged over $4-8$ GHz for various star formation rates as a function of redshift and the respective $3\sigma$ upper limits of our targets. The UV- and radio-derived SFRs for our four targets are summarized in Table \ref{tab:GRB_SFR} along with the stellar masses and ratios of radio-based (total) and UV-based (dust-unobscured) SFRs. 

\subsection{Late-time afterglow emission}

The GRBs have long-lived radio afterglows. Therefore, any estimates of SFR using the radio emission can only be made after the afterglow has faded considerably to ensure minimal/no contamination due to the afterglow. We compiled the past early-time radio observations of the afterglows of our target GRBs available in the literature and extrapolated the afterglow decay using a canonical long GRB radio light curve model (forward shock model) with a $t^{-1}$ decay \citep{chandra2012radio} as follows:

\begin{equation}
  f(t)=\begin{cases}
    F_{m}t_{m}^{-1/2}t^{1/2}, & \text{if $t<t_{m}$}.\\
    F_{m}(t/t_m)^{-1}, & \text{if $t > t_{m}$ }.
  \end{cases}
\end{equation}

\begin{figure*}
\centering
\includegraphics[width=1\textwidth]{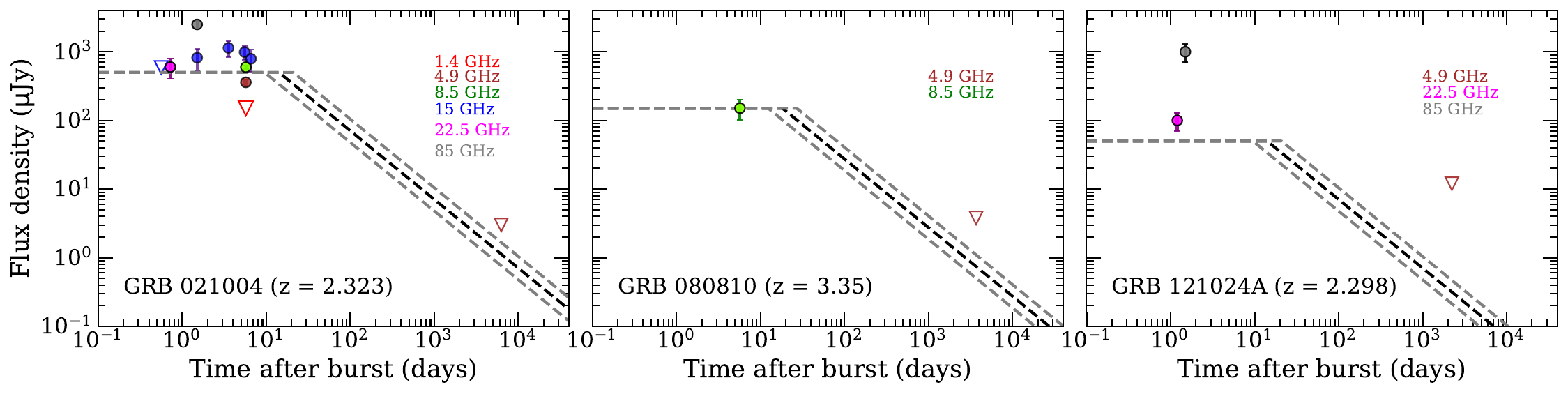}
\figcaption{\label{fig:GRB_Radio_evolution} 
Radio evolution of the afterglows of GRB 021004, GRB 080810, and GRB 121024A, extrapolated using the canonical afterglow evolution model described in Section \ref{sec:radio_SFR}. }
\end{figure*}

Here, $F_{m}$ is the peak flux density at a given frequency and $t_{m}$ is the time of the peak in that frequency. For this extrapolation, we used a conservative approach. We use the latest flux density measurement in C-band (if available) as the peak flux density. If it is not available (eg: GRB 121024A), we extrapolate the flux density using the standard GRB radio afterglow model described in \cite{chandra2012radio}. The typical values of $t_{m}$ range between rest-frame 3 and 6 days at a rest-frame frequency of $\sim$ 25 GHz (which we probe since our targets are at $z \sim 2-3.5$). We translate this $t_{m}$ to the observer frame for each GRB and plot the radio afterglow evolution in Figure \ref{fig:GRB_Radio_evolution}. The three lines show the decay with $t_{m}$ = 3, 4.5, and 6 days (in the rest frame). No early-time radio observations are available for GRB 080310. The conservative approach used here gives the upper limit of radio flux density due to the afterglow and further shows that the late-time radio fluxes for our observations are dominated by the host galaxy and are not likely to be contaminated by the afterglow.

\subsection{SFR in each GRB host}
\begin{figure}
\centering
\includegraphics[width=0.45\textwidth]{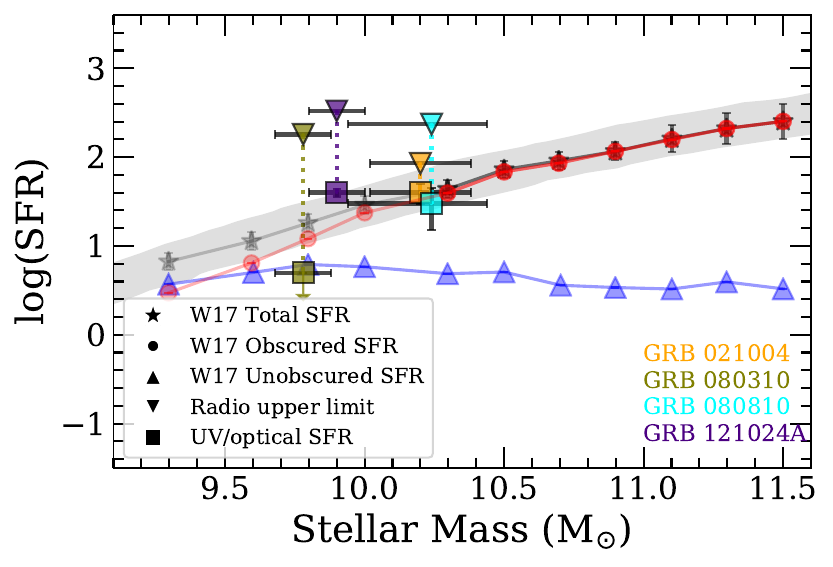}
\figcaption{\label{fig:SFR_vs_Mstar} 
The SFR $-$ $M_{\star}$ relation decomposed into total (black star), obscured (red circle), and unobscured components (blue triangle) of the star formation rate for the galaxies in the CANDELS survey at $z \sim 2-2.5$ \citep{whitaker2017constant}. The gray
band corresponds to the typical 0.3 dex width of the observed relation. Individual GRBs in our sample are shown in various colors with their UV-derived SFR (tracing the dust-unobscured SFR) and the radio-derived SFR (tracing the total SFR).}
\end{figure}

We summarize the UV-derived and radio-derived SFRs for the four GRBs in the following subsections. Note that the $\mathrm{SFR_{unobscured}}$ signifies the uncorrected, UV-based SFR derived from the rest-frame UV luminosity. 
Using the VLA observations, we obtain an estimate of the total SFR ($\mathrm{SFR_{total}}$), independent of assumptions on the dust extinction (in the line of sight or otherwise). 
\newline We also compare the observed ratio  $\mathrm{SFR_{total}/SFR_{unobscured}}$ for our GRB hosts with the same ratio for star-forming galaxies with a similar stellar mass at a redshift range $z \sim 2 - 2.5$, as derived from the CANDELS survey \citep{whitaker2017constant} and summarize this in Figure \ref{fig:SFR_vs_Mstar}.

\subsubsection{GRB 021004}

GRB 021004 is one of the best studied GRBs from the gamma-rays to radio wavelengths. The optical afterglow was detected 3.2 minutes after the prompt high-energy emission and was followed up extensively \citep{fynbo2005afterglow}. The extremely blue host galaxy of GRB 021004 was identified and studied through late-time imaging in the rest-frame UV and optical bands. HST ACS imaging in the F606W band revealed that the host galaxy has a very compact core with a half-light radius of only 0.4 kpc (at $z$ = 2.323).  Based on HST ACS imaging in F606W filter (rest-frame UV),  the impact parameter of the afterglow position is only 0.015\arcsec, corresponding to a distance of 119 pc, which is one of the smallest for long GRBs \citep{fynbo2005afterglow, fruchter2006long, blanchard2016offset}. While this could be a chance projection, it is likely that the GRB progenitor could be associated with a circumnuclear starburst. We note a caveat here that given the typical irregular morphologies of low-mass high-z galaxies, we cannot rule out the small offset being due to the presence of a bright star-forming knot in the rest-frame UV. The line-of-sight extinction $A_{V}$ is 0.20 $\pm$ 0.02 mag (using the SMC extinction law) as derived after 1 week of afterglow decay \citep{fynbo2005afterglow}. The Ly$\alpha$-derived neutral hydrogen column density ($\mathrm{N_{HI}}$) along the line of sight is modest ($\sim$ 10$^{19}$ cm$^{-2}$; \citealt{prochaska2008survey}).

\cite{castro2010grb} derived the host SFR of 40 $\mathrm{M_{\odot} yr^{-1}}$ (without any dust correction) by attributing all of the H$\alpha$ emission to star formation. Given the small $A_{V}$, the dust correction was assumed to be minimal from the afterglow SED. On the other hand, \cite{jakobsson2005ly+} have estimated a lower limit of SFR as 10.6 $\mathrm{M_{\odot} yr^{-1}}$ by converting the Ly$\alpha$ flux to SFR \citep{kennicutt1998star} and assuming a 100\% Ly$\alpha$ escape fraction.  

We derive a 3$\sigma$ upper limit on the C-band flux density of 3.0 $\mu$Jy, corresponding to a radio SFR limit of 85 $\mathrm{M_{\odot} yr^{-1}}$ at $z \sim 2.323$. This result is consistent with the low $A_{V}$ derived from the optical-NIR SED and therefore suggests that the host galaxy as a whole is not significantly affected by dust. This observation identifies a galaxy that is able to sustain a SFR of $\sim$ 40 $\mathrm{M_{\odot} yr^{-1}}$ at $z \sim 2.3$ without significant dust obscuration. Using the non-extinction-corrected H$\alpha$ emission, we get $\mathrm{SFR_{unobscured}}$ = 40 $\mathrm{M_{\odot} yr^{-1}}$, so the ratio $\mathrm{SFR_{total}/SFR_{unobscured}}$ is $< 2.1$ for this $\mathrm{M_{*} > 10^{10} M_{\odot}}$ galaxy. In contrast, the corresponding ratio derived for the main sequence of star-forming galaxies at $z \sim 2.5$ from \cite{whitaker2017constant} is $\sim 6$. 
  
Given the small impact parameter of the afterglow ($119$ pc) from the centroid of the bright star-forming region, the apparent lack of significant dust extinction along the line of sight to the GRB, and in the host galaxy as a whole from the radio observations, is puzzling.   

\subsubsection{GRB 080310}
The afterglow of GRB 080310 was detected 1.5 minutes after the prompt high-energy emission and was followed up extensively \citep[see][for a full discussion]{littlejohns2012origin}. The redshift of this GRB is 2.427 \citep{prochaska2008grb, vreeswijk2008grb}.  \cite{perley2008grb} estimated a low line-of-sight extinction $A_{V}$ of 0.10 $\pm$ 0.05 mag.\ using an SMC-like extinction law (at an average time of $t_0$ $+$ 1750 s). The line-of-sight $\mathrm{N_{HI}}$ is modest ($\sim$ 10$^{18.8}$ cm$^{-2}$). 

The late-time host galaxy imaging using the Low Resolution Imaging Spectrometer (LRIS) on the Keck-I telescope yielded a non-detection with a g-band limiting magnitude of 27.0 \citep{perley2009host}. We estimate a SFR upper limit of 4.5 $\mathrm{M_{\odot} yr^{-1}}$ using the UV luminosity-SFR relation for GRB host galaxies described in \cite{savaglio2009galaxy}. \cite{perley2016swift_2} estimated log($\mathrm{M_{*}/M_{\odot}}$) = 9.8 $\pm$ 0.1 using {\em Spitzer} 3.6 $\mu m$ imaging. However, we caution the reader of the possibility that the {\em Spitzer} 3.6 $\mu m$ flux is contaminated by the diffraction spike from a nearby star despite careful modeling and subtraction of the spike \citep{perley2016swift_2}. 

The VLA observations constrain the SFR to less than 180 $\mathrm{M_{\odot} yr^{-1}}$ (3-$\sigma$ upper limit). However, this limit is not sufficiently deep to constrain the dust obscuration in the host galaxy of GRB 080310.

\subsubsection{GRB 080810}
This is the highest-redshift GRB in our sample at $z = 3.35$. The afterglow of GRB 080810 was detected 80 seconds after the prompt emission by the X-ray telescope (XRT; \cite{burrows2005swift}) and UV-optical telescope (UVOT; \cite{roming2005swift}) on board the Neil Gehrels Swift Observatory \citep{gehrels2004swift}. \cite{prochaska2008grb} obtained the optical spectra of the afterglow using the Keck HIRES spectrograph starting 37 minutes after the trigger and derived a redshift of 3.35. The Ly$\alpha$-derived line-of-sight $\mathrm{N_{HI}}$ is small ($\sim$ 10$^{17.5}$ cm$^{-2}$). We refer the readers to \cite{page2009multiwavelength} for a discussion of the extensive multi-wavelength follow-up of this GRB.   

Extensive late-time ground-based photometry and spectroscopy of the host galaxy of GRB 080810 revealed an extended structure with a bright compact region \citep[see][for more details]{wiseman2017gas}. Further, a strong detection of redshifted Ly$\alpha$ emission at a redshift of 3.36 confirmed the association of the GRB and the detected host galaxy \citep{wiseman2017gas}. They estimate a modest host extinction of $A_{V} \sim$ 0.4 mag.\ from SED fitting. \cite{greiner2015gamma} convert the extinction-corrected UV luminosity to SFR (using the $\mathrm{L_{UV}-SFR}$ relation in \cite{duncan2014mass} and $A_{1600} \sim$ 1.3 mag.) to obtain SFR $\sim$ 100 $\mathrm{M_{\odot} yr^{-1}}$, which is further corroborated by SED fitting \citep{wiseman2017gas}. The uncorrected SFR is $\sim$ 30 $\mathrm{M_{\odot} yr^{-1}}$. The stellar mass, derived from the {\em Spitzer} 3.6 $\mu m$ photometry, is log($\mathrm{M_{*}/M_{\odot}}$) = 10.2 $\pm$ 0.1 \citep{perley2016swift}.

Here we report the first ever deep late-time radio observation of a GRB with a spectroscopic redshift $z > 3.1$. We derive a 3-$\sigma$ upper limit on the C-band flux density of 3.8 $\mu$Jy, corresponding to a radio-based SFR upper limit of 235 $\mathrm{M_{\odot}~yr^{-1}}$ at $z \sim 3.35$. The dust-corrected SFR from the UV SED is therefore consistent with the total SFR limit derived from the radio observations. This further implies that the modest $A_{V}$ estimated from the UV SED fitting reasonably takes into account the dust correction. 

Using the uncorrected UV SFR, we derive a ratio $\mathrm{SFR_{total}/SFR_{unobscured}}$ $< 7$ for this $\mathrm{M_{*} > 10^{10}~M_{\odot}}$ galaxy. This is consistent with the corresponding ratio derived for the main sequence of star-forming galaxies at $z \sim 2-2.5$ from \cite{whitaker2017constant}, which  gives $\mathrm{SFR_{total}/SFR_{unobscured}} \sim 6$. 
Here, we extrapolate the non-evolution of this ratio from $z \sim 2.5$ to 3.3 for the star-forming galaxies on the main sequence at a given $\mathrm{M_{*}}$, as presented in \cite{whitaker2017constant}. 

\subsubsection{GRB 121024A}

The afterglow of GRB 121024A was followed up 93 seconds after the prompt emission by the X-ray telescope (XRT; \citealt{burrows2005swift}) on board the Neil Gehrels Swift Observatory \citep{gehrels2004swift}.    
\cite{tanvir2012grb} obtained the optical/NIR spectra of the afterglow using the X-shooter spectrograph on the Very Large Telescope (VLT) and determined a redshift of 2.298. The line-of-sight $\mathrm{N_{HI}}$ of 10$^{21.5}$ cm$^{-2}$ indicates that this is a damped Ly$\alpha$ system. We refer the readers to \cite{friis2015warm} for a detailed summary of the extensive multi-wavelength follow-up of this GRB.

Various emission lines including H$\alpha$, H$\beta$, [O II] $\lambda\lambda$3727, 3729 doublet, [N II] $\lambda$6583, and [O III] $\lambda\lambda$4959, 5007 were detected in the X-shooter NIR spectrum of the afterglow. Extensive optical and NIR photometry of the host galaxy was obtained using VLT/HAWK-I, NOT, and GTC \citep[see][for details]{friis2015warm}. The stellar population synthesis modelling of the host yielded a modest extinction $A_{V}$ of 0.15 $\pm$ 0.15 mag.\ and $\mathrm{log(M_{*}/M_{\odot})}$ = 9.9$^{+0.2}_{-0.3}$. 

\cite{friis2015warm} estimate the SFR from the extinction-corrected H$\alpha$ and [O II] fluxes as 42 $\pm$ 11 and 53 $\pm$ 15 $\mathrm{M_{\odot}~yr^{-1}}$ using conversion factors from \cite{kennicutt1998star}. However, note that the extinction correction to the SFR is small ($\sim 15\%$). They further corroborate this SFR by stellar population synthesis modelling. 

The 3-$\sigma$ upper limit on the C-band flux density of GRB 121024A is 18 $\mu$Jy. The relatively higher background is due to a bright source at 6$^\prime$ angular separation. Using the VLA observations, we obtain a 3-$\sigma$ upper limit of the total SFR as 500 $\mathrm{M_{\odot} yr^{-1}}$. However, this limit is not sufficiently deep to constrain the dust obscuration in the host galaxy of GRB 121024A. The limiting $\mathrm{SFR_{total}/SFR_{unobscured}} < 12.5$ is consistent with the corresponding expected ratio ($\sim 5$) from \cite{whitaker2017constant} for a star-forming galaxy of this stellar mass on the main sequence at $z \sim 2-2.5$.    

\begin{deluxetable*}{ccccccc}[t]
\tablecaption{Summary of GRB host properties \label{tab:GRB_SFR}}
\tablehead{
\colhead{GRB\tablenotemark{a}} &
\colhead{$z$} &
\colhead{\makecell{$\mathrm{log(N_{HI})}$\tablenotemark{a} \\ (cm$^2$) }} &
\colhead{\makecell{M$_*$\\ ($\mathrm{M_{\odot}}$)}} &
\colhead{\makecell{SFR$\mathrm{_{UV}}$\\ ($\mathrm{M_{\odot}~yr^{-1}}$)}} & 
\colhead{\makecell{SFR$\mathrm{_{Radio}}$\tablenotemark{b}\\ ($\mathrm{M_{\odot}~yr^{-1}}$)}} & 
\colhead{$\mathrm{\frac{SFR_{total}}{SFR_{UV}}}$} }
\startdata
021004 & 2.323 & 19.00 $\pm$ 0.2\tablenotemark{c} & 10.2 $\pm$ 0.18\tablenotemark{g} & 40 $\pm$ 10 & $<$ 85 & $<$ 2.1 \\
080310 & 2.427 & 18.80 $\pm$ 0.1\tablenotemark{d} & 9.78 $\pm$ 0.2\tablenotemark{h} & $<$ 5 & $<$ 180 & $-$  \\
080810 & 3.35 & 17.5 $\pm$ 0.15\tablenotemark{e} & 10.24$\pm$ 0.1\tablenotemark{h}& 30 $\pm$ 15 & $<$ 235 & $<$ 7.8 \\
121024A & 2.298 & 21.5 $\pm$ 0.1\tablenotemark{f} & 9.9$^{+0.2}_{-0.3}$\tablenotemark{f} & 40 $\pm$ 4 &  $<$ 330 & $<$ 8.3\\
\enddata
\vspace{3ex}
$^{a}${Ly$\alpha$-derived $\mathrm{N_{HI}}$}
$^{b}${3$\sigma$ upper limit}, $^{c}${\cite{prochaska2008survey}}, $^{d}${\cite{fox2008high}}, $^{e}${\cite{page2009multiwavelength}}, $^{f}${\cite{friis2015warm}}, $^{g}${\cite{savaglio2009galaxy}}, $^{h}${\cite{perley2016swift_2}}
\end{deluxetable*}

\section{Discussion} \label{sec: discussion}
\begin{figure}
\centering
\includegraphics[width=0.5\textwidth]{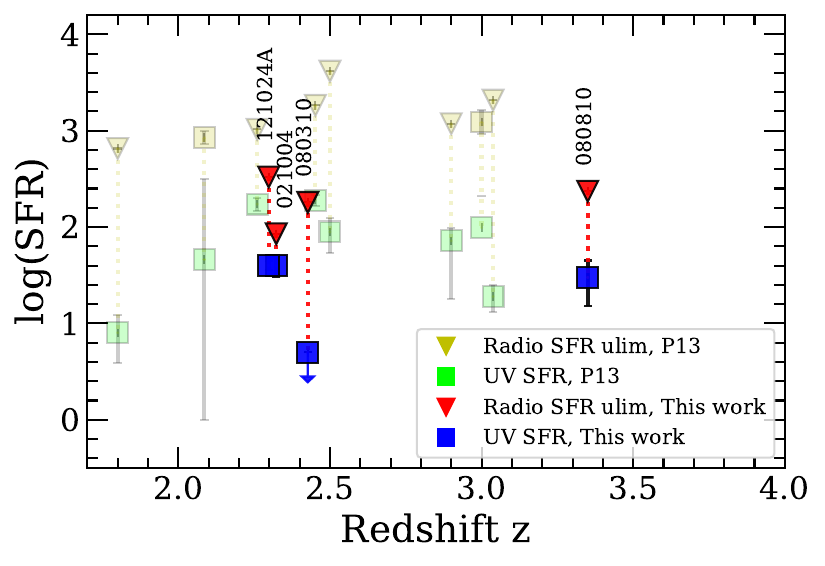}
\figcaption{\label{fig:SFR_vs_z_lit} 
The comparison of the radio-derived SFR (tracing the total SFR) and UV-derived SFR (tracing the dust-unobscured SFR) as a function of redshift for the four GRBs presented here (in the foreground), and GRBs in the literature in the background. P13: \cite{perley2013population} and one data point (GRB 060814) from \cite{greiner2016probing}.}
\end{figure}

\noindent
Our observations have targeted massive ($\mathrm{M_{*} > 10^{9.5}~M_{\odot}}$) high-z GRBs ($z \sim 2 - 3.5$) with high-resolution and high SNR rest-frame UV afterglow spectra (i.e.\ a rest-frame UV-bright afterglow). 
 Previous studies have observed the host galaxies of so-called `dark' GRBs (rest-frame UV/optically dark afterglows) in radio \citep{perley2013radio, perley2015connecting, greiner2016probing}. These observations are summarized in Figure \ref{fig:SFR_vs_z_lit}.  
However, we caution the readers that in Figure \ref{fig:SFR_vs_z_lit}, the UV-based SFR from the literature are dust-corrected. In the future, a combined sample of the radio observations for the hosts of GRBs with UV-bright afterglows and UV-dark afterglows can help address the question as to whether GRB hosts are biased against the highly dust-obscured starburst population at high redshifts. This question has strong implications for the use of GRBs as SFR tracers at high redshift. Deeper radio limits (comparable to this paper) for the dark GRB hosts will be needed to address this question in the future.



The radio flux limits in our observations are a least 3 times deeper than the previous limits on the SFR at $z > 2$ \citep{perley2015connecting}, and thus provide tighter constraints on whether GRB hosts at these redshifts are more likely to be dusty starburst galaxies or not. Out of the four GRBs in this sample, we have well-defined upper limits of the $\mathrm{SFR_{Total}/SFR_{UV}}$ for three of them (see Table \ref{tab:GRB_SFR}). We compare these limits with the observed distribution of the dust-obscuration ratios at high redshifts from the CANDELS survey in \cite{whitaker2017constant} (hereafter W17; see Figure 2 therein; we compare against the inverse of  $1 - \mathrm{f_{obscured}}$).  

The upper limits of $\mathrm{SFR_{Total}/SFR_{UV}}$ for the host galaxies of GRB 080810 ($<$ 7.8) and 121024A ($< 8.3$) are consistent with this distribution. We note here that the upper limits are derived using 3-$\sigma$ radio flux limits. 
On the other hand, for GRB 021004, the $\mathrm{SFR_{Total}/SFR_{UV}}$ $\lesssim$ $2$. Only 1\% of the W17 sample with the corresponding stellar mass ( $\mathrm{log(M_{*}/M_{\odot}) = 10.2 \pm 0.2}$) falls in the $\mathrm{SFR_{Total}/SFR_{UV}}$ $\lesssim$ $2$ regime. Hence, GRB 021004 is inconsistent with being drawn from the W17 distribution. The ratio for GRB 080310 is unconstrained due to UV non-detection. 

Given that 50\% of our sample limits are consistent with the results of W17 (using 3-$\sigma$ limits), 25\% of the sample is  inconsistent with W17, and 25\% is unconstrained, we can only draw a coarse conclusion. The deep radio limits suggest that the overall star formation activity in these GRB hosts is not heavily obscured by dust (i.e. $\mathrm{SFR_{Total}/SFR_{UV}} < 10$, unlike LIRGs; \citealt{bouwens2009uv, howell2010great, casey2014dusty}), and possibly slightly less obscured than the star-forming main sequence population at $z \sim 2.5$ \citep{speagle2014highly}.

Particularly, GRB 021004 provides a striking example of lack of significant dust obscuration in the central region of a star-forming galaxy at $z > 2$, given that the separation of the GRB from the galaxy centroid is only 119 pc \citep{fynbo2005afterglow, fruchter2006long}. The sightline extinction, derived from the afterglow is also small ($\mathrm{A_{V}}$ = $0.2 \pm 0.02$ mag.). Two possible scenarios can explain these results: a) 
the GRB occurred in a locally dusty cloud but globally, the host galaxy lacks significant amount of dust. The low sightline extinction would then imply that the burst 
occurred along a clear sightline within its star-forming cloud. b) the GRB occurred in a star-forming region which has cleared the dust from past star formation and the overall galaxy also lacks significant amount of dust. The GRB sightline would then be a representative sightline.  



The results from our limited sample suggest that the GRBs with UV-bright afterglows (i.e. optically thin sighltines in UV) at $z \sim 2-3.5$  are likely to be star-forming galaxies with SFRs moderately higher ($< 5 \times$) than the star-forming main sequence \citep{speagle2014highly}, but without significant dust obscuration in their star-forming regions. 
 
However, it is likely that this result only applies to the GRBs with UV-bright afterglows due to our selection criteria. At the same time, the dust extinction along a sightline may not necessarily represent the dust obscuration on a galaxy scale, for optically thin as well as optically thick sightlines (in UV). 
More radio observations of GRB hosts at $z  > 2$ with a depth at least $2 \times \mathrm{SFR_{UV}}$  are necessary to confirm this hypothesis. This is required for GRBs with UV-bright afterglows as well as with UV/optically dark afterglows to rule out any selection bias based on the line-of-sight extinction. 

\section{Summary} \label{sec:summary}
\noindent 


If the GRBs are unbiased tracers of star formation at high redshifts ($z > 2$), then we should expect that a large fraction of GRB hosts are highly dust-obscured starbursting galaxies, since these are well known to be major contributors to the cosmic star formation at high redshifts. 
The goal of our study was to investigate the galaxy-scale dust obscuration in the GRB hosts with optically thin sightlines in the UV.
We conducted deep radio observations of a 
subset of four massive ($\mathrm{M_{*} > 10^{9.5}}$ $\mathrm{M_{\odot}}$) GRB hosts at $z > 2$ for which high signal-to-noise (typical SNR $\sim$ 10) and high-resolution ($\Delta v$ $<$ 50 km s$^{-1}$) rest-frame UV spectra of the afterglow are available. The selected targets are GRB 021004, GRB 080310, GRB 080810, and GRB 121024A. We measured the total SFR (= obscured $+$ unobscured SFR) of the hosts using VLA C-band observations and compared them against the unobscured component of the SFR, measured from the rest-frame UV luminosity. The depth of the radio observations in this study has allowed us to put tight constraints on the ratio of the total-to-unobscured SFRs ($\mathrm{SFR_{total}/SFR_{unobscured}}$). 

We find that the  radio-based star formation rates are in general not substantially higher than those obtained from the optical/UV measurements. Thus, the fraction of total star formation that is obscured by dust ($\mathrm{SFR_{obscured}/SFR_{total}}$) in most of the GRB hosts, even at $z > 2$, is less than 90\%, unlike LIRGs or dusty starburst galaxies.  Particularly, for the well-constrained case of GRB 021004 ($z = 2.323$), we find that the upper limit of the radio-based `total SFR' is less than twice the UV-based `unobscured SFR' of the GRB hosts (thus, $\mathrm{SFR_{obscured}/SFR_{total}} < 50\%$). Our results suggest that the 
dust obscuration in the star-forming regions of these galaxies is small, and sometimes (e.g.\ for GRB 021004) even smaller than the dust obscuration seen in typical main-sequence star-forming galaxies at these redshifts.  We reiterate that the results obtained here may only apply to GRBs with UV-bright afterglows.

The present upper limits on the radio-based SFRs prevent us from determining where the GRB host population lies with respect to the main sequence of star-forming galaxies at $z > 2$. Deeper radio observations to a depth of $2 \times \mathrm{SFR_{UV}}$ are required to answer this question. Currently, we are limited by the sensitivity of the radio instrumentation (eg: JVLA) to reach these deep limits. They will be achievable with the higher sensitivity of  upcoming radio telescope arrays such as ng-VLA and SKA1-MID.


\vspace*{3ex}

P.G. was supported by NASA Earth and Space Science Fellowship (ASTRO18F-0085) for this research. The authors are grateful to Drs.\ K. Whitaker, S. Bradley Cenko, and Daniel Perley for their useful comments in the early stages of this paper. We thank Dr.\ Ashley Zauderer,  Nicholas Ferraro, and Virginia Cunningham for helpful discussions about data analysis. The National Radio Astronomy Observatory is a facility of the National Science Foundation operated under cooperative agreement by Associated Universities, Inc. 

\vspace{2mm}
\facilities{VLA}


\software{{\fontfamily{qcr}\selectfont CASA} \citep{mcmullin2007casa}, {\fontfamily{astropy}\selectfont astropy} \citep{robitaille2013astropy}  }

\bibliographystyle{apj}
\bibliography{ref}

\end{document}